\documentclass[a4paper]{article}

\usepackage{INTERSPEECH2021}

\title{Towards Multi-Scale Style Control for Expressive Speech Synthesis}
\name{Xiang Li$^{1,\dagger}$, Changhe Song$^{1,\dagger}$, Jingbei Li$^1$, Zhiyong Wu$^{1,2,*}$, \thanks{$\dagger$ Equal contribution. * Corresponding author.} Jia Jia$^{1,3}$, Helen Meng$^{1,2}$}

\address{
  $^1$Tsinghua-CUHK Joint Research Center for Media Sciences, Technologies and Systems,\\Shenzhen International Graduate School, Tsinghua University, Shenzhen, China\\
   $^2$Department of Systems Engineering and Engineering Management,\\The Chinese University of Hong Kong, Hong Kong SAR, China \\
   $^3$Department of Computer Science and Technology, Tsinghua University, Beijing, China}
\email{\{xiang-li20, sch19, lijb19\}@mails.tsinghua.edu.cn, \{zywu, hmmeng\}@se.cuhk.edu.hk, jjia@tsinghua.edu.cn}

\begin{document}

\maketitle

\begin{abstract}
This paper introduces a multi-scale speech style modeling method for end-to-end expressive speech synthesis. 
The proposed method employs a multi-scale reference encoder to extract both the global-scale utterance-level and the local-scale quasi-phoneme-level style features of the target speech, which are then fed into the speech synthesis model as an extension to the input phoneme sequence. 
During training time, the multi-scale style model could be jointly trained with the speech synthesis model in an end-to-end fashion. 
By applying the proposed method to style transfer task, experimental results indicate that the controllability of the multi-scale speech style model and the expressiveness of the synthesized speech are greatly improved. 
Moreover, by assigning different reference speeches to extraction of style on each scale, the flexibility of the proposed method is further revealed.
  
\end{abstract}


\section{Introduction}

\begin{figure*}[t]
  \centering
  \includegraphics[width=0.8\linewidth]{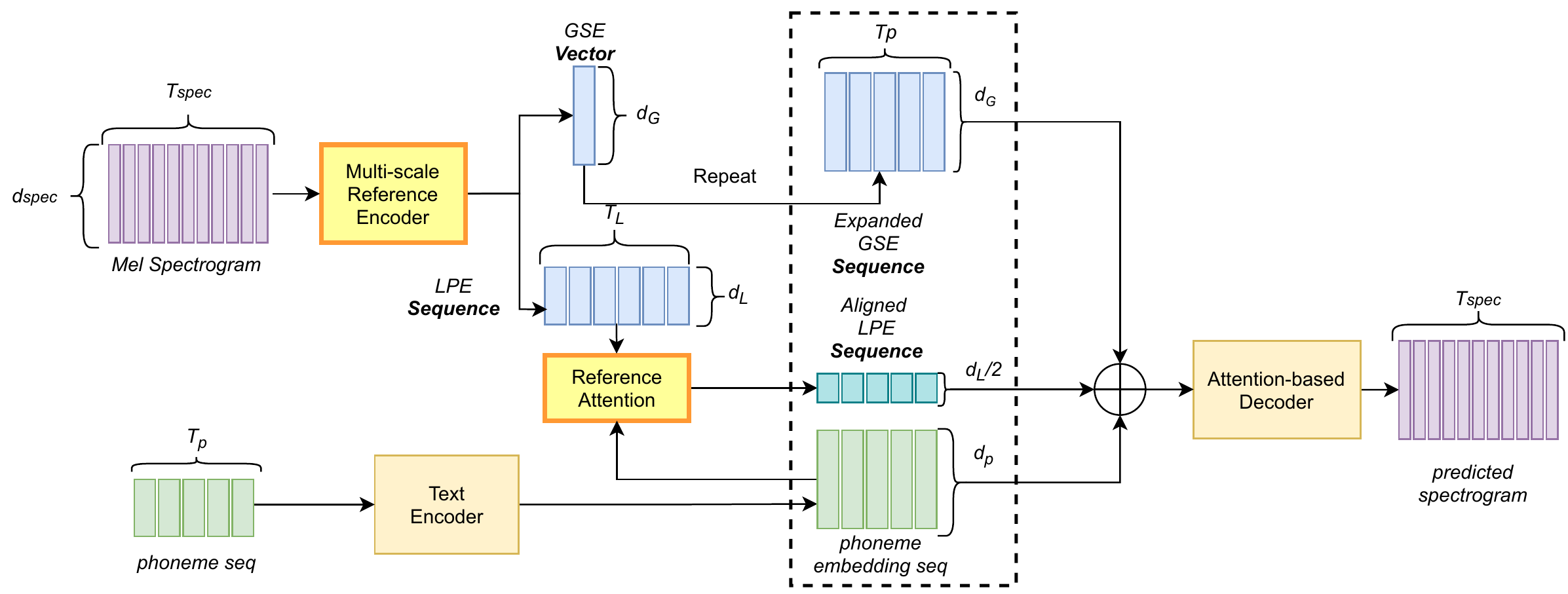}
  \caption{Schematic diagram of the proposed model}
  \label{fig:scheme}
\end{figure*}

Due to the prosperous development of deep learning, statistical parametric speech synthesis models built with artificial neural networks have achieved profound advancement on both the quality and naturalness of the synthesized speech \cite{6639215, qian2014training}. 
Popular end-to-end text-to-speech (TTS) models like Tacotron are now embodied with the ability to generate almost human-like audios with the support of high fidelity neural vocoder models \cite{Wang2017, shen2018natural}. 
But the limited expressiveness of synthesized audio persists as one of the major gaps between synthesized speech and real human speech, which draws growing attention to expressive speech synthesis studies.

Frequently mentioned expressive speech synthesis methods like reference encoder introduce the idea of extracting a global style embedding from the given reference audio \cite{skerry2018towards}. 
Global style token and their derivatives further manage to memorize and reproduce the global-scale style features of speech \cite{wang2018style, wu2019end}.
Other more recently proposed methods seek to replace the global-scale latent style embedding in previous works with fine-grained latent prosody embedding sequence.
\cite{lee2019robust} incorporates an additional reference attention mechanism to aligned the extracted prosody embedding sequence to phoneme sequence.
Other researches like \cite{Klimkov2019} achieve the same goal by resorting to force-alignment tools.
\cite{daxin2020fine} further proposes phoneme-scale content-independent style extracting module based on force-alignment.

However, most of these existing expressive speech synthesis methods focus on only style features of one single scale.
On one hand,
reference encoder and global style tokens accomplish the modeling of global-scale speech style, but suffer from the absence of local-scale style modeling, and are thus incapable of controlling the fine-grained prosody of the synthesized speech. 
On the other hand,
fine-grained models usually find it quite challenging to recognize the utterance-level style context due to the removal of global-scale style features, and are consequently not provided with the ability to control the global style of the synthesized speech directly.

Meanwhile, the actual expressiveness of human speech can be perceived as a compound of multi-scale acoustic factors. 
One is the global-scale speech style, which includes, but not limited to the timbre and emotion of the speaker.
Style of this level are supposed to be consistent throughout one single utterance.
The other is the local-scale speech style, which comprises the speed, energy, pitch, pause and other acoustic features of the speech. 
Fluctuation of these features is frequently observed even among different locations within the same utterance.

Some latest researches are observed to devote effort to performing a multi-scale style modeling on some specific tasks like emotional speech synthesis.
For example, \cite{lei2021fine} employs a learnt ranking function \cite{zhu2019controlling} to extract local-scale emotion intensity scalars of each phoneme in the utterance.
The extracted emotion intensity sequence and a global-scale emotion label are then together fed into the TTS system in order to generate speech with desired strength fluctuation, as well as matching the target emotion category.
But in these methods, style modules of different scales are not accommodated into the end-to-end train process of the whole TTS system, or heavily relies on auxiliary labels. 
This implies that the local-scale style modules are not aware of the utterance-level style context, and the learnt global-scale latent style embedding might eventually reach an average of different samples with the same label, rather than matching the global-scale style features of given utterance specifically.

With all the listed imperfections of previous works taken into consideration, this paper proposes a multi-scale speech style modeling method for end-to-end expressive speech synthesis, which is composed by a multi-scale reference encoder, a reference attention mechanism and a vanilla encoder-decoder-based TTS backbone.

The multi-scale reference encoder is trained to extract a global-scale latent style embedding (GSE) vector and a local-scale latent prosody embedding (LPE) sequence from the mel-spectrogram of the given reference speech.
And following \cite{lee2019robust}, a reference attention is assigned to learn the alignment between the local-scale prosody embedding sequence and the phoneme sequence. 
Both the multi-scale reference encoder and the reference attention are jointly trained with the TTS system in an end-to-end fashion.
Moreover, instead of inheriting the frame-level granularity from the input spectrogram, downsampling is introduced in our multi-scale reference encoder to obtain intermediate latent features with the granularity closer to human spoken phoneme.

Benefiting from these design features, our method is able to control the global-scale and local-scale speech styles simultaneously by assigning reference audios.
We prove the advantages of the proposed method by conducting style transfer experiment on an emotional dataset.
Through additional cross-emotion and non-parallel transfer experiments, the flexibility and potentiality of our model is further revealed upon a closer inspection.


\section{Methodology}




Our multi-scale style modeling method consists of two major components:
(i) A multi-scale reference encoder that extracts the 
global-scale style embedding (GSE) vector 
and the local-scale prosody embedding (LPE) sequence 
from the mel-spectrogram of the given reference speech;
(ii) A reference attention mechanism which aligns the extracted 
LPE sequence
to the phoneme embedding sequence produced by the text encoder.
Based on above components,
an end-to-end encoder-decoder based TTS backbone is constructed to achieve controllability and improved expressiveness.

\subsection{Multi-scale expressive TTS system}

As shown in Figure \ref{fig:scheme}, given an input reference spectrogram with $d_{spec}$ bands and $T_{spec}$ frames, the multi-scale reference encoder generates a GSE vector with dimensionality of $d_G$, as well as an LPE sequence with dimensionality of $d_L$ and length of $T_L$.
The reference attention, which aims to align the LPE sequence to the phoneme embedding sequence, outputs an aligned LPE sequence.
On the other hand, the GSE vector is repeated $T_p$ times along time dimension into an extended GSE sequence $(d_G\times T_p)$.
With the same length $T_p$, the aligned LPE sequence and the expanded GSE sequence are concatenated altogether with the phoneme embedding sequence, which produces a $(d_p+d_L/2+d_G)\times T_p$ latent feature sequence further fed to the attention-based decoder.

\begin{figure}[t]
  \centering
  \includegraphics[width=\linewidth]{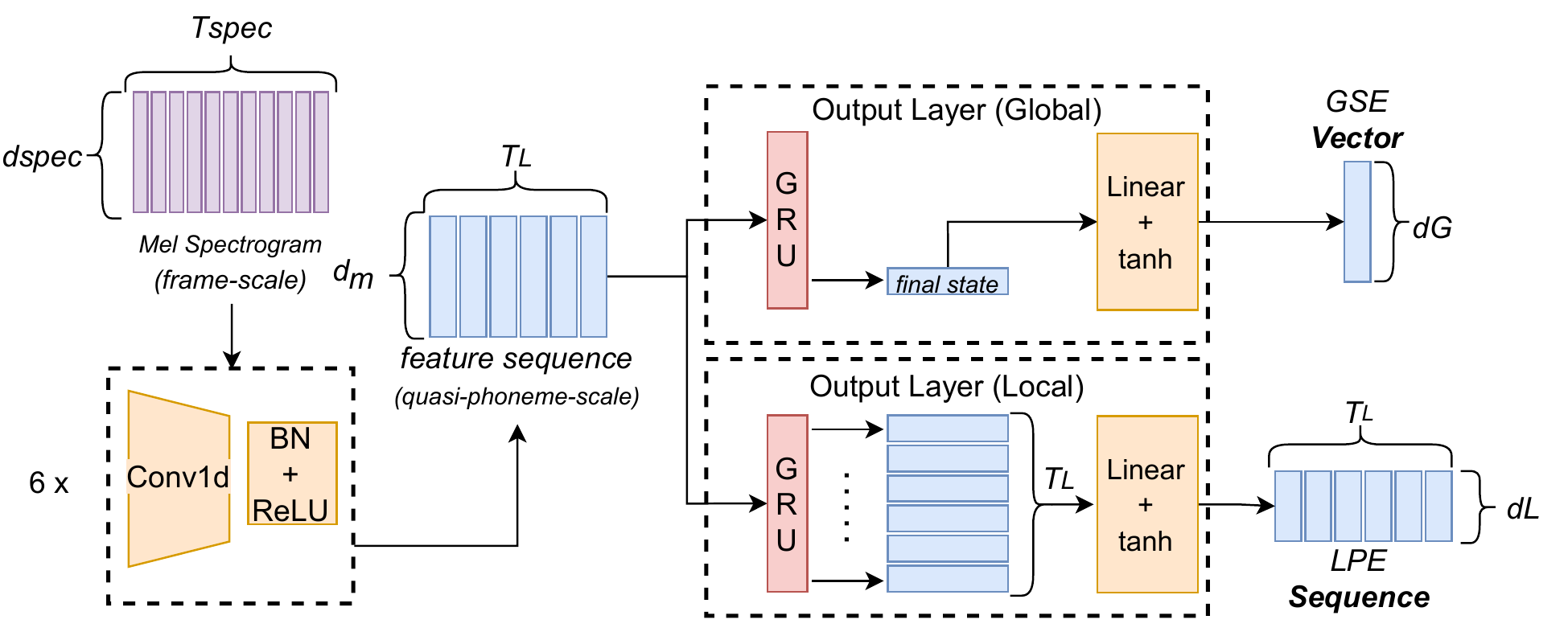}
  \caption{Flowchart of the Multi-scale Reference Encoder}
  \label{fig:ms_ref_enc}
\end{figure}


\begin{table*}[th]
    \caption{MOS Evaluation Result}
    \label{tab:mos}
    \centering
    \begin{tabular}{l|ccc|ccc}
        \toprule
         & \multicolumn{3}{c|}{Emotion (Global-scale style)} & \multicolumn{3}{c}{Fine-grained Prosody (Local-scale style)} \\
         & Proposed & Base-G & Base-L & Proposed & Base-G & Base-L \\
        \midrule
         Overall & $\mathbf{4.155\pm0.552}$ & $4.149\pm0.510$ & $2.955\pm1.569$ & $\mathbf{4.136\pm0.701}$ & $3.471\pm1.406$ & $3.625\pm0.901$ \\
        \midrule
         Neutral & $\mathbf{4.425\pm0.328}$ & $4.308\pm0.463$ & $2.742\pm1.875$ & $\mathbf{4.375\pm0.418}$ & $3.958\pm0.890$ & $3.825\pm0.661$ \\
         Angry & $3.883\pm0.836$ & $\mathbf{4.192\pm0.538}$ & $3.200\pm0.993$ & $\mathbf{4.075\pm0.836}$ & $3.583\pm1.393$ & $3.558\pm1.047$ \\
         Fear & $\mathbf{4.225\pm0.458}$ & $4.167\pm0.372$ & $3.767\pm0.929$ & $\mathbf{3.975\pm0.974}$ & $3.275\pm1.233$ & $3.575\pm1.028$ \\
         Disgust & $\mathbf{3.942\pm0.688}$ & $3.892\pm0.680$ & $2.325\pm1.136$ & $\mathbf{3.983\pm0.733}$ & $3.350\pm1.577$ & $3.542\pm0.882$ \\
         Happy & $\mathbf{4.300\pm0.360}$ & $4.200\pm0.427$ & $2.192\pm1.305$ & $\mathbf{4.208\pm0.532}$ & $3.200\pm1.327$ & $3.475\pm0.983$ \\
         Sad & $\mathbf{4.267\pm0.396}$ & $4.133\pm0.482$ & $3.833\pm0.739$ & $\mathbf{4.200\pm0.677}$ & $3.492\pm1.417$ & $3.758\pm0.800$ \\
         Surprised & $4.042\pm0.557$ & $\mathbf{4.150\pm0.511}$ & $2.625\pm1.384$ & $\mathbf{4.133\pm0.616}$ & $3.442\pm1.630$ & $3.642\pm0.813$ \\
        \bottomrule
    \end{tabular}
\end{table*}

\subsection{Multi-scale reference encoder}
To extract both global and local scale style features from the given reference speech, we introduce a specifically designed multi-scale reference encoder,
which is made up by a stack of 6 convolution layers and 2 output layers, as shown in Figure \ref{fig:ms_ref_enc}.

\subsubsection{1D-Convolution downsampling layers}

For the convolution layers, 1D-convolution along temporal dimension is adopted to help the reference attention to learn the alignment between the output local-scale prosody embedding sequence and the phoneme embedding sequence.
Each of the convolution layers is composed by 3x1 filters, ReLU activation and batch normalization \cite{ioffe2015batch}.

In addition, in order to regulate the temporal granularity of the convolution output closer to human vocal perception, 
the filter strides of 6 convolution layers are set as [2, 1, 2, 1, 2, 2].
Through this, the input spectrogram $(d_{spec}\times T_{spec})$ is downsampled to $(d_m\times T_L)$, where $d_m$ is the number of filters in the last convolution layer, and $T_L$ is equal to $\lceil T_{spec}/16\rceil$. 
For any input spectrogram with a common frame shift like 12.5 msec, the temporal granularity of the outcome features is approximately 200 msec, 
which is in line with
the common duration range of consonant-vowel syllables
from 150 msec to 200 msec \cite{steinschneider2013representation}.
The downsampling operation ensures that after the convolution stack, the temporal granularity of the intermediate feature sequence is properly reformed to a quasi-phoneme-scale.
In this way, the robustness of reference attention is enhanced.

\subsubsection{Scale-specific output layers}
Output layers of local-scale and global-scale share the same structure, which is made up of consecutive Gated Recurrent Unit (GRU) \cite{DBLP:conf/emnlp/ChoMGBBSB14} and fully-connected layer, as well as tanh activation. 
Both output layers take the above quasi-phoneme-scale feature sequence $(d_m\times T_L)$ as input.
However, inside the global-scale output layer, only the final state of GRU is fed to subsequent layers. 
The outcome of global-scale output layer is forced to be a latent style vector $(d_G \times 1)$, while that of local-scale output layer is still a quasi-phoneme-scale prosody embedding sequence $(d_L\times T_L)$.

\subsection{Reference attention mechanism}
In terms of expressive speech synthesis, fine-grained style embedding sequence is usually reformed into sequence with the same length as phoneme sequence, in order that it can be used as an extension by the speech synthesis backbone.

In our model, scaled dot-product attention \cite{NIPS2017_3f5ee243} is employed as the reference attention mechanism, which learns to find the alignment between the quasi-phoneme-scale LPE sequence $(d_L\times T_L)$ from the reference encoder and the target phoneme embedding sequence.
The quasi-phoneme-scale LPE sequence is splitted along feature dimension into two sub-sequencies $(d_L/2\times T_L)$, which are fed as the key and value of the reference attention mechanism respectively.
Meanwhile, the phoneme embedding sequence is fed as the query.
Finally, the reference attention outputs an aligned LPE sequence $(d_L/2\times T_p)$ with the same length as the phoneme embedding sequence.

\subsection{Model usage}
During training time, the whole end-to-end system are jointly trained. 
And the reference speech is exactly the target speech of the TTS system.
However, in order to prevent the learning of style modeling on different scales from disturbing each other, the training process is divided into 2 stages.
In the first stage, the global-scale output layer of the reference encoder is not inserted into the model so that the local-scale style model and the reference attention is trained without disturbance.
While in the second stage, the training of global-scale output layer is involved, and the parameters of text encoder, reference attention, as well as reference encoder (excluding the global-scale output layer) are frozen.

During inference time, by accepting input text and reference speech, the proposed model can synthesize speech with style close to the reference speech.
Normally, the global and local scale share the same reference speech.
But in our model, unique reference speech for each scale can be accepted respectively.
Here, the reference speech for local scale is often required to have the same content with the input text, while the content of the reference speech for global scale is not restricted.

\section{Experiments}

\subsection{Dataset and model details}

An internal single-speaker emotional corpus on Mandarin is employed in our experiment. The emotions of all speeches are classified by 7 categories (neutral, angry, fear, disgust, happy, sad, surprised). A total of 28.79 hours speech data are formed as 22,500 utterances, among which 10,500 belong to the neutral emotion, and the other part is evenly divided by the rest emotions. Additionally, about 15\% of the whole dataset is parallel: for any text content in this part, there are 7 speech recordings corresponding to 7 different emotions respectively.

For the backbone of our expressive speech synthesis system, Tacotron 2 \cite{shen2018natural} with decoder reduction factor set to 3 is adopted.
In the mutli-scale reference encoder, the number of GRU hidden units in both global and local output layer is set to 128.
And in terms of dimensionalities of multi-scale style features, $d_L$ is set to 6 as information bottleneck \cite{lee2019robust}, while $d_G$ is kept as 128.
Following the common training trick on emotional dataset, a vanilla emotion classifier composed of 2 linear layers is inserted to obtain more discriminative GSE vectors,
which is jointly trained with the rest of the model by cross-entropy loss.
As for waveform generation, HiFi-GAN \cite{DBLP:conf/nips/KongKB20} is chosen as our neural vocoder.


\subsection{Mutli-scale style transfer}


\subsubsection{Ablation study}

In order to demonstrate the influence of GSE and LPE in our proposed multi-scale speech style modeling method, two mono-scale style models are established as our baselines.

\textbf{Base-G}: The global-scale baseline model, which shares the same TTS backbone and multi-scale reference encoder, but excludes the reference attention mechanism and local output layer in reference encoder.

\textbf{Base-L}: The local-scale baseline model, where the global output layer in reference encoder is removed.

\begin{figure}[t]
  \centering
  \includegraphics[width=0.95\linewidth]{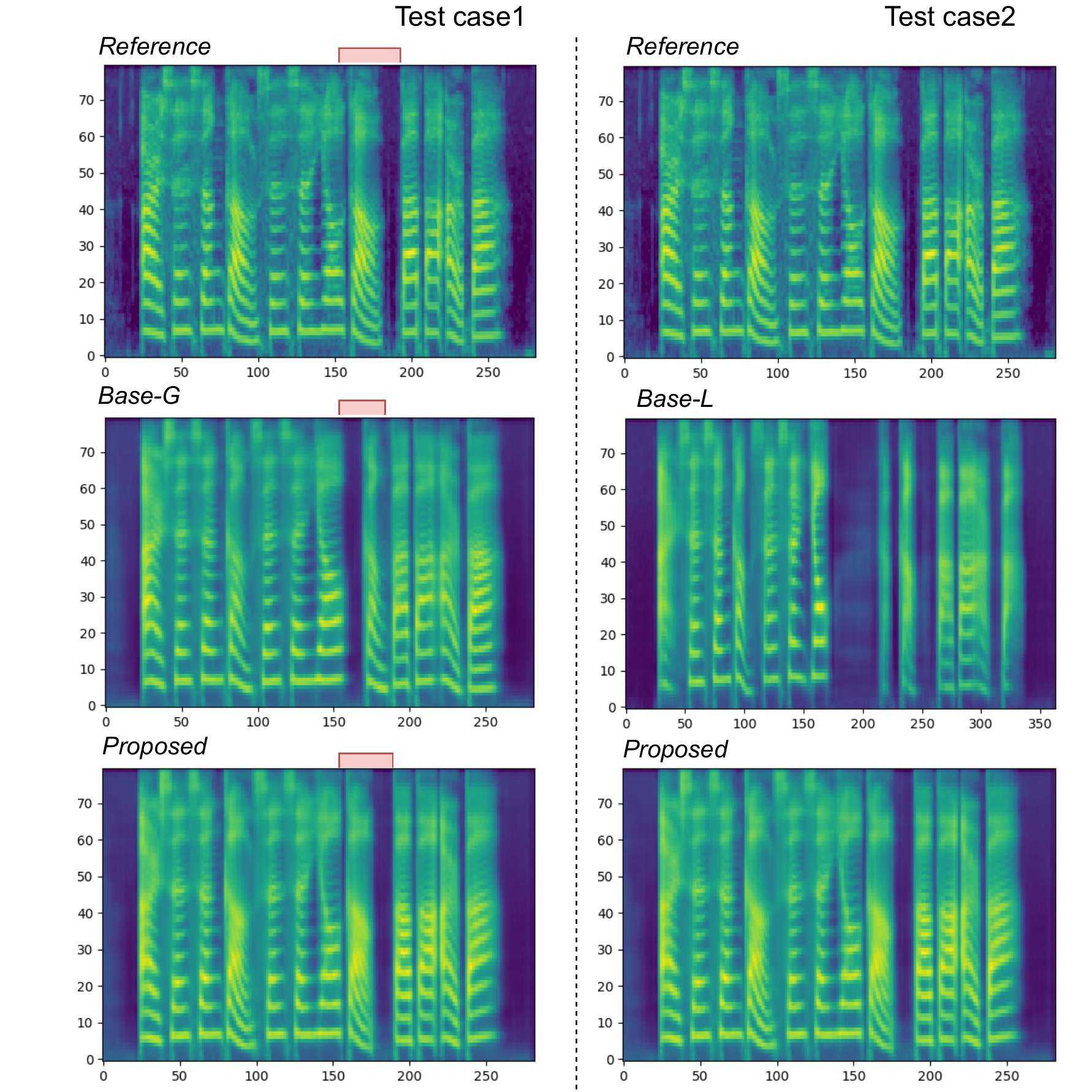}
  \caption{Comparison on parallel transfer result. (Upper row: reference speech, middle row: speech produced by the Base-G (Test case 1) / Base-L (Test case 2) model, bottom row: speech produced by the proposed model.)}
  \label{fig:case1}
\end{figure}

The style modeling abilities of the proposed model and the two baseline models are investigated by applying them to parallel style transfer task, where each of the tested model is given a reference audio with the same content as input text, to generate speech with style as close to the reference as possible.

In Figure \ref{fig:case1}, the 8-th character in the reference speech of test case 1 is followed by a pause (which are noted by the red rectangles).
The pause in the speech generated by Base-G is mistakenly located in front of the 8-th character, revealing that the Base-G model is troubled with inaccurate transfer result on fined-grain speech prosody.
On the other hand, as depicted in test case 2 of Figure \ref{fig:case1}, the formant characteristic of mel-spectrogram predicted by Base-L model is apparently different from the give reference speech.
Which indicates that due to the absence of global style context, the global emotion styles of speeches generated by the Base-L model fail to match the original emotion.
While none of these mistakes described above are observed in the speech generated by our proposed multi-scale model.
And the enhancement on the similarity between synthesized speech and reference speech also shows the promoted controllability of the proposed model.

\begin{figure}[t]
  \centering
  \includegraphics[width=\linewidth]{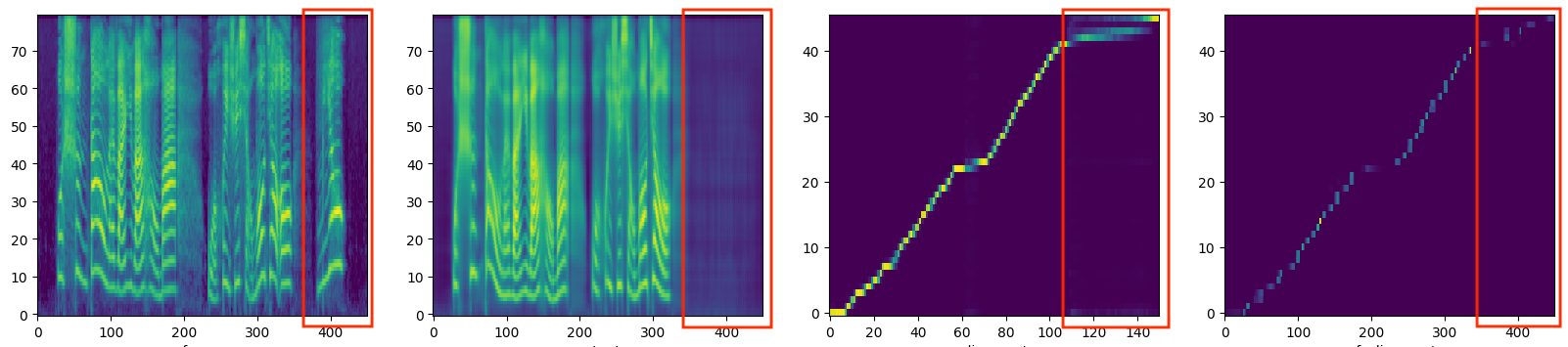}
  \includegraphics[width=\linewidth]{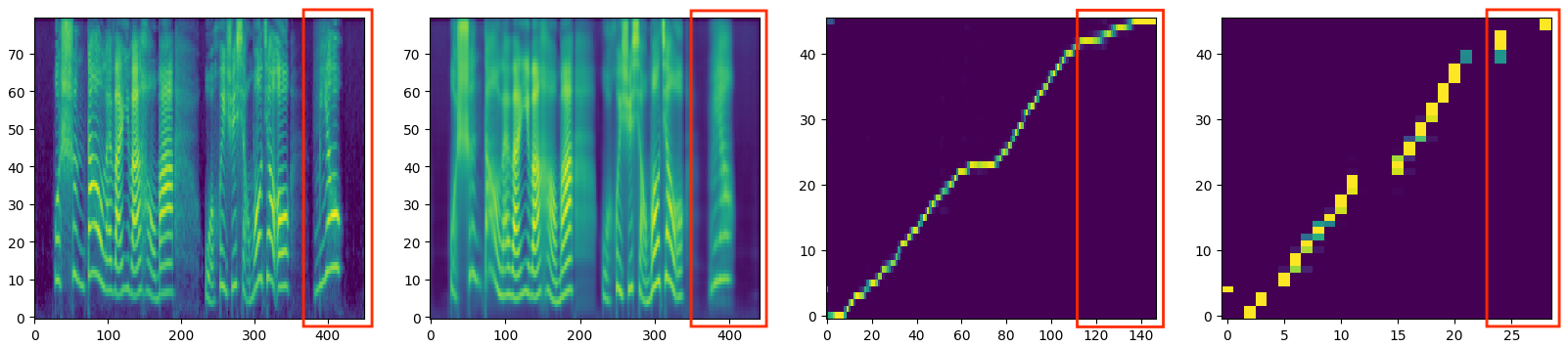}
  \caption{Comparison on the robustness of reference attention (Upper row: test result on Base-FS model, lower row: test result on proposed quasi-phoneme-scale model. From left to right, each column is: reference speech, predicted speech, alignment of decoder attention, alignment of reference attention.)}
  \label{fig:case2}
\end{figure}

\subsubsection{Subjective evaluation result}

The described advantages of multi-scale style model is further verified by conducting Mean Opinion Score (MOS) test on the parallel style transfer result of the proposed and baseline models.
An amount of 10 test reference audio groups are established, each of which is made up of 7 speech audios with the same text content but unique emotion category and various prosody. 
The participants are asked to give two subjective scores for the similarity of the synthesized speech to the reference speech, in terms of global emotion and local prosody respectively.

As presented in Table \ref{tab:mos}, the proposed model outperforms the baselines on the precision of both global-scale and local-scale style transfer.
Between the baseline models, the global-scale Base-G model possesses noticeable outperformance on global emotion transfer, while the speech produced by local-scale Base-L model tends to be better controlled on reproducing fined-grained prosody.
This indicates that since each mono-scale baseline style model possesses its own expertise in style modeling on the corresponding scale, by integrating them into a whole multi-scale style model, significant improvement on the overall performance can be achieved.

\subsection{Quasi-phoneme-scale v.s. Frame-scale}
To demonstrate the impact of regulating the temporal granularity of style feature to quasi-phoneme-scale, we trained another multi-scale model, setting the strides of all convolution layers in the reference encoder to 1 (noted as \textbf{Base-FS}).
Thus, the temporal granularity of the intermediate feature sequence in the reference encoder is rolled back to frame scale,  as well as the extracted LSE sequence.

In Figure \ref{fig:case2}, given the same reference audio, Base-FS fails to synthesize the final character in the sentence (encompassed by the red rectangles), while the proposed quasi-phoneme-scale model is not encountered with such problem.
And the alignment of the reference attention in Base-FS is evidently less bright in the area of the last character, which arouses confusion in the decoding process, and eventually damages the quality of synthesized speech.
These phenomena reveal the robustness of the reference attention received considerable degradation due to the removal of downsampling.

\subsection{Multi-reference style transfer}


Different from previous mono-scale style models which only support accepting one single reference speech,
our multi-scale model is innately adaptable to multiple reference speeches with considerable diversities.
As long as the text content of the local reference is correspondent with the input text, arbitrary global reference speech could be adopted to transfer the embedded global-scale style to the synthesized speech, regardless of its local-scale style and content.


Experiment evidences have confirmed that,
the transfer result preserves the local-scale prosody factors of the local-scale reference, while its global-scale emotion style is altered to match the global-scale reference.
Since the effectiveness of global emotion transfer typically varies among the subjective perception of each and every individual, checking our demo page\footnote{https://thuhcsi.github.io/interspeech2021-multi-scale-style-control} for detailed samples is strongly recommended.

\section{Conclusions}
In this work, a novel multi-scale style modeling method for expressive speech synthesis is introduced.
Experimental results support that the designed multi-scale scheme improves the effectiveness of style modeling on both global and local scales.
Further attempts on cross-emotion and partially non-parallel style transfer brings an insight to the practical flexibility of our method, which indicates the significance of the proposed model towards multi-scale style control of the generated speech.



\newpage

\bibliographystyle{IEEEtran}

\bibliography{bibtext}


\end{document}